\documentclass[12pt,aps,prd,preprint,tightenlines,superscriptaddress,showpacs,nofootinbib]
{revtex4-1}
\usepackage{epsfig}
\usepackage{amssymb,amsmath}
\usepackage{color}
\usepackage[utf8]{inputenc} 

\newcommand{\bea}{\begin{eqnarray}}
\newcommand{\eea}{\end{eqnarray}}

\begin{document}
%
\vspace*{1.0cm}

\begin{center}
\baselineskip 20pt 
{\Large\bf 
Light $Z^\prime$ and Dark Matter from U(1)$_X$ Gauge Symmetry
}
\vspace{1cm}

{\large 
Nobuchika Okada $^1$, Satomi Okada $^1$ and Qaisar Shafi $^2$ 
}
\vspace{.5cm}

{\baselineskip 20pt \it
$^1$Department of Physics and Astronomy, University of Alabama, Tuscaloosa, AL35487, USA \\
$^{2}$ Bartol Research Institute, Department of Physics and Astronomy, \\
University of Delaware, Newark, DE 19716, USA
} 

\vspace{.5cm}

\vspace{1.5cm} {\bf Abstract}
\end{center}

We consider a U(1)$_X$ gauge symmetry extension of the Standard Model (SM) 
   with a $Z^\prime$-portal Majorana fermion dark matter that allows for a relatively light gauge boson 
   $Z^\prime$ with mass of 10 MeV$-$ a few GeV and a much heavier dark matter
   through the freeze-in mechanism. 
In a second scenario the roles are reversed, and the dark matter mass, in the keV range or so, 
   lies well below the $Z^\prime$ mass, say, $\sim 1$ GeV. 
We outline the parameter space that can be explored for these two scenarios 
   at the future Lifetime Frontier experiments including Belle-II, FASER, LDMX and SHiP.

\thispagestyle{empty}

\newpage

\addtocounter{page}{-1}

\baselineskip 18pt

\section{Introduction} 

The minimal $B-L$ (Baryon number minus Lepton number) model 
   \cite{Pati:1973uk, Davidson:1978pm, Mohapatra:1980qe, Marshak:1979fm,
     Wetterich:1981bx, Masiero:1982fi, Mohapatra:1982xz, Buchmuller:1991ce} 
  is a simple and well-motivated extension of the Standard Model (SM)     
  to supplement the SM with a neutrino mass generation mechanism. 
In the model, the accidental global U(1)$_{B-L}$ symmetry in the SM is gauged and 
   the SM particle content is extended to include three right-handed neutrinos (RHNs) and 
   an SM-singlet $B-L$ Higgs field with 2 units of $B-L$ charge. 
All the anomalies associated with the $B-L$ gauge symmetry are canceled in the presence of three RHNs. 
The $B-L$ symmetry is broken by a vacuum expectation value (VEV) of the $B-L$ Higgs field, 
   generating the $B-L$ gauge boson ($Z^\prime$) mass as well as Majorana masses for the three RHNs. 
With the RHNs Majorana masses, 
   the desired light neutrino masses are naturally induced through the seesaw mechanism
  \cite{Minkowski:1977sc, Yanagida:1979as, GellMann:1980vs, Mohapatra:1979ia}.

A concise way to supplement the minimal $B-L$ model with a DM candidate has been proposed in Ref.~\cite{Okada:2010wd} 
   where, instead of introducing a new particle for a DM candidate, 
   a discrete $Z_2$ symmetry (parity) is introduced while keeping the minimal $B-L$ model particle content intact.\footnote{
Note that this $Z_2$-parity is quite distinct from the $Z_2$ subgroup of U(1)$_{B-L}$ 
   that is left unbroken by the VEV of the $B-L$ Higgs field. 
   In the framework of SO(10) this latter $Z_2$ is a subgroup of $Z_4$, 
   the center of SO(10) \cite{Kibble:1982ae}. 
}
\footnote{
For a different approach to dark matter and gauge boson in a U(1) setting, 
see Ref.~\cite{Fortes:2017kca} and references therein.
}    
Under this $Z_2$-parity, one RHN ($N_R$) is assigned an odd parity, while all the other fields in the model are even. 
This parity assignment ensures the stability of $N_R$, which is a unique DM candidate in the model. 
It is known that only two RHNs are sufficient for reproducing the neutrino oscillation data 
   and this setup is called ``minimal seesaw" \cite{King:1999mb, Frampton:2002qc}. 
The $Z_2$-parity categorizes the three RHNs into $2+1$ RHNs: two RHNs for the minimal seesaw mechanism 
   and one RHN for the DM candidate.

The phenomenology of Majorana fermion DM $N_R$ has been extensively studied.  
This DM particle can communicate with the SM particles through the Higgs boson exchange (Higgs-portal) 
  and/or the $Z^\prime$ boson exchange ($Z^\prime$-portal). 
A detailed analysis for the Higgs-portal case is found in Refs.~\cite{Okada:2010wd, Okada:2012sg, Basak:2013cga}, 
  and the $Z^\prime$-portal case is discussed in Refs.~\cite{Okada:2016gsh, Escudero:2018fwn} 
  (see Ref.~\cite{Okada:2018ktp} for a review).  
The $Z^\prime$-portal case is particularly interesting since it has been pointed out that the DM physics and 
   the search for a $Z^\prime$ boson resonance at the Large Hadron Collider (LHC) complement each other 
   in narrowing down the allowed model-parameter space.   
For a similar study of a $Z^\prime$-portal Dirac fermion DM in the context of the minimal $B-L$ model, 
   see Refs.~\cite{FileviezPerez:2018toq, FileviezPerez:2019cyn}.

In most of previous studies, the RHN DM has sizable couplings with the SM particles 
  and it had been in thermal equilibrium in the early universe. 
In such a thermal DM scenario, the DM relic density in the present universe is determined 
  by the freeze-out mechanism, namely, the DM particle decouples from the thermal plasma
  of the SM particles at some point in the early universe,
  and the yield (co-moving number density) of the DM particle is essentially fixed at the freeze-out time.  
However, in general, we can also consider a non-thermal DM scenario 
  in which the coupling of a DM particle with the SM particles is too weak to allow it  to be in thermal equilibrium with the SM particles.  
Given the fact that no compelling evidence for a DM particle has so far been observed in the direct
  and indirect DM search experiments, it seems well-motivated to consider the non-thermal DM scenario. 
In this case, the DM relic density is determined by the ``freeze-in'' mechanism \cite{McDonald:2001vt, Hall:2009bx} 
  (see Ref.~\cite{Bernal:2017kxu} for a review), namely, the DM particles are produced 
  from the thermal plasma of the SM particles, 
   assuming a vanishing initial DM density at reheating after inflation. 
The freeze-in DM scenario has recently attracted a fair amount of attention.

For the RHN DM in the minimal $B-L$ model, the freeze-in DM scenario has been studied 
  in Ref.~\cite{Kaneta:2016vkq, Biswas:2016bfo}.
In particular, the authors in Ref.~\cite{Kaneta:2016vkq} have studied in detail the $Z^\prime$-portal RHN DM 
  with a light $Z^\prime$ boson. 
It has been pointed out that a parameter space which reproduces the observed DM density 
   through the freeze-in mechanism predicts the light $Z^\prime$ boson to be long-lived, 
   and its displaced vertex signature can be searched for by the planned/proposed experiments at the Lifetime Frontier.   
The results in Ref.~\cite{Kaneta:2016vkq} for the freeze-in RHN DM 
 are complementary to those in Ref.~\cite{Okada:2016gsh} for the thermal RHN DM.

It is straightforward to generalize the minimal $B-L$ model to the so-called minimal U(1)$_X$ model, 
   where the U(1)$_X$ charge of a field is assigned as a linear combination of its SM hyper-charge
   and $B-L$ charge with a new real parameter $x_H$ \cite{Appelquist:2002mw, Oda:2015gna}.  
Except for the generalization of the U(1)$_X$ charge assignments, the particle content of the model
   is the same as that of the minimal $B-L$ model. 
All of the U(1)$_X$ related anomalies are canceled in the presence of three RHNs. 
This is because the SM hyper-charge and the $B-L$ gauge interactions are anomaly-free
   in the minimal $B-L$ model, and the U(1)$_X$ charge for each chiral fermion is defined as their linear combination. 
The $Z^\prime$-portal RHN DM scenario in the context of the minimal U(1)$_X$ model 
   has been investigated in Ref.~\cite{Okada:2016tci} (see also Refs.~\cite{Oda:2017kwl, Okada:2017dqs}).  
It has been found that complementarity between the thermal RHN DM phenomenology and the LHC search
   for the $Z^\prime$ boson narrows down the allowed parameter region.

Motivated by the future Lifetime Frontier experiments, we investigate in this paper the freeze-in RHN DM in the minimal U(1)$_X$ model. 
One may regard our study as a completion of the previous work in Ref.~\cite{Okada:2016tci}
   for the thermal RHN DM. 
We consider a relatively light $Z^\prime$ boson with a mass in the range of 10 MeV$-$a few GeV, 
  which can be explored by the future Lifetime Frontier experiments. 
As for the DM mass, we consider two possibilities:
  Case (i) has a DM particle that is much heavier than the $Z^\prime$ boson, 
  and in Case (ii) the DM particle is much lighter with a mass of ${\cal O}$(keV).  
For both cases, we calculate the DM relic density at present and identify the parameter region 
  that reproduces the observed DM density and can be explored by the future Lifetime Frontier experiments.

This paper is organized as follows: 
In the next section, we define the minimal U(1)$_X$ model with the $Z^\prime$-portal RHN DM. 
In Sec.~\ref{sec:3}, we investigate the case with a very small U(1)$_X$ gauge coupling, 
   so that the RHN DM had never been in thermal equilibrium with the SM particles.  
We calculate the DM relic abundance through the freeze-in mechanism and 
   identify the parameter region to reproduce the observed DM density. 
In Sec.~\ref{sec:3-1} the RHN DM mass ($m_{DM}$) is set to be much larger 
   than the $Z^\prime$ boson mass ($m_{Z^\prime}$), and 
   we identify the allowed parameter regions for various values of $x_H$.  
We will see an impact of $x_H$ values on the search for a long-lived $Z^\prime$ boson
   at the future Lifetime Frontier experiments. 
The case $m_{DM} < m_{Z^\prime}$ is analyzed in Sec.~\ref{sec:3-2}. 
Sec.~\ref{sec:4} is devoted to conclusion and discussion.

\section{$Z^\prime$-portal RHN DM in the minimal U(1)$_X$ model }
\label{sec:2}
\begin{table}[t]
\begin{center}
\begin{tabular}{|c|ccc|c|c|}
\hline
      &  SU(3)$_c$  & SU(2)$_L$ & U(1)$_Y$ & U(1)$_X$  & $ Z_2 $\\ 
\hline \hline
$q^{i}_{L}$ & {\bf 3 }    &  {\bf 2}         & $ \frac{1}{6} $       & $\frac{1}{6} x_{H} + \frac{1}{3} $    & $+$\\
$u^{i}_{R}$ & {\bf 3 }    &  {\bf 1}         & $ \frac{2}{3}$       & $  \frac{2}{3} x_{H} +  \frac{1}{3}$   & $+$\\
$d^{i}_{R}$ & {\bf 3 }    &  {\bf 1}         & $- \frac{1}{3}$       & $-\frac{1}{3} x_{H} +   \frac{1}{3}$   &$+$\\
\hline
$\ell^{i}_{L}$ & {\bf 1 }    &  {\bf 2}         & $- \frac{1}{2}$       & $-\frac{1}{2} x_{H} + (-1) $  & $+$ \\
$e^{i}_{R}$    & {\bf 1 }    &  {\bf 1}         & $-1$                   & $ -x_{H} +(-1) $   & $+$ \\
\hline
$H$            & {\bf 1 }    &  {\bf 2}         & $- \frac{1}{2}$       & $ -\frac{1}{2} x_{H}$  & $+$ \\  
\hline
$N^{j}_{R}$    & {\bf 1 }    &  {\bf 1}         &$0$                    & $- 1$   & $+$ \\
$N_{R}$         & {\bf 1 }    &  {\bf 1}         &$0$                    & $- 1$   & $-$ \\
$\Phi$            & {\bf 1 }       &  {\bf 1}       &$ 0$                  & $ + 2$  & $+$ \\ 
\hline
\end{tabular}
\end{center}
\caption{
The particle content of the minimal U(1)$_X$ model with $Z_2$ symmetry (parity). 
In addition to the SM particle content ($i=1,2,3$), three RHNs ($N_R^j$ ($j=1,2$) and $N_R$) 
  and the U(1)$_X$ Higgs field ($\Phi$) are introduced. 
Due to its $Z_2$-parity assignment, the $N_R$ is a unique DM candidate. 
The U(1)$_X$ charge of a field is defined as a linear combination of its U(1)$_Y$ and U(1)$_{B-L}$ charges 
  with a real parameter $x_H$. 
The minimal $B-L$ model is defined as a limit of $x_H \to 0$. 
}
\label{table1}
\end{table}

The particle content of our model is listed in Table~\ref{table1}. 
The U(1)$_X$ charge of a particle is defined as a linear combination of its U(1)$_Y$ and U(1)$_{B-L}$ charges 
  with a real parameter $x_H$. 
Note that the minimal $B-L$ model is realized by setting $x_H=0$,  
  while the U(1)$_X$ gauge interaction becomes similar (up to a sign) to the SM hyper-charge 
  interaction for $|x_H| \gg 1$ (``hyper-charge oriented'' U(1)$_X$ \cite{Okada:2017cvy}). 
All the gauge and mixed gauge-gravitational anomalies are canceled by the presence of three RHNs. 
The parity-odd $N_R$ is stable and a unique DM candidate (RHN DM) in the model.

The U(1)$_X$ charge assignments in Table~\ref{table1} allow all the SM Yukawa couplings for quarks 
  and charged leptons. 
In addition, the following gauge invariant Yukawa couplings are introduced: 
\bea
\mathcal{L}_{Y} \supset  - \sum_{i=1}^{3} \sum_{j=1}^{2} Y^{ij}_{D} \overline{\ell^i_{L}} H N_R^j 
          -\frac{1}{2} \sum_{k=1}^{2} Y^k_N \Phi \overline{N_R^{k~C}} N_R^k 
          -\frac{1}{2}  Y_N \Phi \overline{N_R^{~C}} N_R 
  + {\rm h.c.} ,
\label{Lag1} 
\eea
where the first term is the neutrino Dirac Yukawa coupling, and the second and 
   third terms are the Majorana Yukawa couplings for RHNs. 
Without loss of generality, we work in the basis where the Majorana Yukawa coupling matrix is already diagonalized.  
Note that due to $Z_2$-parity, only two RHNs ($N_R^{j=1, 2}$) are involved 
  in the neutrino Dirac Yukawa coupling. 
Associated with the U(1)$_X$ gauge symmetry breaking by a nonzero VEV of  $\Phi$, 
  the RHNs acquire Majorana masses.  
The minimal seesaw mechanism with only two RHNs operates after electroweak symmetry breaking, 
  and the desired light neutrino masses are generated naturally.  
Even with two RHNs, the Yukawa sector has a sufficient number of free parameters in $Y_D^{ij}$ 
   for reproducing the neutrino oscillation data and predicting one massless neutrino eigenstate.  
The observed baryon asymmetry in the Universe can also be reproduced through leptogensis \cite{Fukugita:1986hr}, 
   with only two RHNs~\cite{Frampton:2002qc}
   (see, for example, Refs.~\cite{Blanchet:2009bu, Iso:2010mv, Dev:2017xry} for detailed analysis of leptogenesis at the TeV scale
   in the presence of the $B-L$ gauge interaction).

We introduce the following scalar potential for the SM Higgs doublet ($H$) and the U(1)$_X$ Higgs field ($\Phi$):  
\bea  
V = \lambda_H \left(  H^{\dagger}H -\frac{v^2}{2} \right)^2
+ \lambda_{\Phi} \left(  \Phi^{\dagger} \Phi -\frac{v_X^2}{2}  \right)^2
+ \lambda_{\rm mix} 
\left(  H^{\dagger}H -\frac{v^2}{2} \right) 
\left(  \Phi^{\dagger} \Phi -\frac{v_X^2}{2}  \right) , 
\label{Higgs_Potential }
\eea
where all quartic couplings are chosen to be positive. 
At the potential minimum, the Higgs fields develop their respective VEVs:  
\bea
  \langle H \rangle =  \left(  \begin{array}{c}  
    \frac{v}{\sqrt{2}} \\
    0 \end{array}
\right),  \; \,  \; 
\langle \Phi \rangle =  \frac{v_X}{\sqrt{2}}. 
\eea
In this paper, we assume $\lambda_{\rm mix} \ll1$ and neglect the mass mixing 
   between the SM Higgs boson and the U(1)$_X$ Higgs boson. 
In this case, the RHN DM communicates with the SM particles only through the $Z^\prime$ boson ($Z^\prime$-portal RHN DM).  
Through the Higgs VEVs, the Majorana masses of RHNs and $Z^\prime$ boson mass are expressed as 
\bea 
  m_N^j=\frac{Y_N^j}{\sqrt{2}} v_X,  \; \; 
  m_{DM}=\frac{Y_N}{\sqrt{2}} v_X,  \; \; 
  m_{Z^\prime} = g_X \sqrt{4 v_X^2+  \frac{v^2}{4}} \simeq 2 g_X v_X , 
\label{Mass}
\eea  
where $g_X$ is the U(1)$_X$ gauge coupling, and we have assumed $v_X^2 \gg v^2$. 
This hierarchy between the two VEVs is required by the LEP constraint \cite{Carena:2004xs, Heeck:2014zfa}. 
As we will see below, this hierarchy is required for our freeze-in DM scenario 
    with a light $Z^\prime$ boson that we focus on in this paper.

\section{Freeze-in RHN DM}
\label{sec:3}
As has been investigated in Refs.~\cite{Okada:2016gsh, Okada:2016tci}, 
  the thermal RHN DM can be viable for a limited parameter space, 
  namely, $m_{DM} \simeq m_{Z^\prime}/2$ and $m_{Z^\prime}={\cal O}$(TeV). 
This is because (i) the LHC constraints from the search for a $Z^\prime$ boson resonance   
  are very severe and the U(1) gauge coupling is restricted to be small 
  (for example, $g_X \sim 0.01- 0.1$ for $m_{Z^\prime} =2$ TeV (see Ref.~\cite{Das:2019fee})), 
  and (ii), with such a small gauge coupling, an enhancement of the DM annihilation cross section 
  through a $Z^\prime$ boson resonance is necessary to reproduce the observed DM relic density.  
On the other hand, for the freeze-in RHN DM that we focus on in this paper, 
  $g_X \ll 0.01$ and both DM particle and $Z^\prime$ boson can be light.  
Here we first give general formulas that we employ in our analysis.

As is well-known for a thermal DM scenario, the DM relic density is evaluated by solving the Boltzmann equation \cite{Kolb:1990vq}: 
\bea 
\frac{d Y}{d x}
= -\frac{s(m_{DM})}{H(m_{DM})} \,  \frac{\langle\sigma v_{\rm rel} \rangle}{x^2} \, (Y^2-Y_{EQ}^2) ,
\label{Y:Boltzmann1}
\eea 
where the temperature of the universe is normalized by the mass of the RHN DM as $x=m_{DM}/T$, 
   $H(m_{DM})$ and  $s(m_{DM})$ are the Hubble parameter 
   and the entropy density of the universe at $T=m_{DM}$, respectively, 
   $Y$ is the DM yield (the ratio of the DM number density to the entropy density), 
   $Y_{EQ}$ is the yield of the DM particle in thermal equilibrium, 
  and $\langle \sigma v_{\rm rel} \rangle$ is the thermal average of the DM annihilation cross section 
  times relative velocity ($v_{\rm rel}$). 
Explicit formulas of the quantities involved in the Boltzmann equation are as follows: 
\bea 
s(T)= \frac{2  \pi^2}{45} g_\star T^3 ,  \; \;
H(T) =  \sqrt{\frac{\pi^2}{90} g_\star} \frac{T^2}{M_P},  \; \;
s Y_{EQ}= \frac{g_{DM}}{2 \pi^2} \frac{m_{DM}^3}{x} K_2(x),   
\eea
where $M_P=2.43 \times 10^{18}$  GeV is the reduced Planck mass, 
   $g_{DM}=2$ is the number of degrees of freedom for the Majorana fermion DM, 
   $g_\star$ is the effective total number of degrees of freedom for the particles in thermal equilibrium 
   (in the following analysis, we use $g_\star=106.75$ for the SM particles),  
   and $K_2$ is the modified Bessel function of the second kind.   
The thermal averaged annihilation cross section is given by 
\bea 
\langle \sigma v \rangle = \left(s Y_{EQ} \right)^{-2} \, g_{DM}^2 \,
  \frac{m_{DM}}{64 \pi^4 x} 
  \int_{4 m_{DM}^2}^\infty  ds \,  
  2 (s- 4 m_{DM}^2) \, \sigma(s)  \, \sqrt{s} K_1 \left(\frac{x \sqrt{s}}{m_{DM}}\right) , 
\label{ThAvgSigma}
\eea
where $\sigma(s)$ is the DM pair annihilation cross section, and $K_1$ is the modified Bessel function of the first kind.

Although the freeze-in DM particle was never in thermal equilibrium with the SM particle plasma, 
  we can employ the above Boltzmann equation also for evaluating the freeze-in DM relic density. 
This is because the second term on the right-hand side of Eq.~(\ref{Y:Boltzmann1}) 
  is the DM pair production rate from the annihilations of the SM particles in the thermal plasma. 
In solving the Boltzmann equation, a crucial difference between the thermal DM and freeze-in DM 
  lies in the initial condition. 
For the thermal DM case, we set the initial condition to be $Y(x_{RH})=Y_{EQ}(x_{RH})$ for $x_{RH} \ll 1$ 
  while $Y(x_{RH})=0$ for the freeze-in DM case, 
  where $x_{RH}=m_{DM}/T_{RH}$ with the reheat temperature ($T_{RH}$) after inflation, 
  and we have assumed that the freeze-in DM particle has no direct coupling with the inflaton. 
Solving the Boltzmann equation with a suitable initial condition, the DM relic density at present is evaluated by
\bea
 \Omega_{\rm DM} h^2 = \frac{m_{DM} \, Y(\infty) \, s_0}{\rho_c/h^2}, 
\eea
where $s_0=2890/{\rm cm}^3$ is the entropy density of the present universe, and 
  $\rho_c/h^2=1.05 \times 10^{-5}$ GeV/cm$^3$ is the critical density.  
This must reproduce the observed DM relic density set by the Planck 2018 measurements \cite{Aghanim:2018eyx}: 
$\Omega_{\rm DM} h^2 = 0.12$.

\subsection{Case (i): $m_{Z^\prime} \ll  m_{DM}$}
\label{sec:3-1} 
We first consider the case with $m_{DM} \gg m_{Z^\prime}$ by setting 10 MeV $\lesssim m_{Z^\prime} \lesssim$ 1 GeV, 
   which is the mass range of $Z^\prime$ boson to be explored by the Lifetime Frontier experiments (see below). 
The main process for the DM pair creation from the SM thermal plasma is  $f \bar{f} \to Z^\prime \to NN$ \cite{FileviezPerez:2018toq}, 
   and the corresponding cross section is give by 
\bea 
 \sigma(s)=\frac{g_X^4}{48 \pi}  \frac{\sqrt{s (s-4 m_{DM}^2)}}
  {(s-m_{Z^\prime}^2)^2+m_{Z^\prime}^2 \Gamma_{Z^\prime}^2} 
    F(x_H),  
\label{DMSigma}
\eea 
where $\Gamma_{Z^\prime}$ is the total decay width of $Z^\prime$ boson, and 
\bea 
  F(x_H)=13+ 16 x_H  + 10 x_H^2 
 \label{F}  
\eea 
  if all the SM fermions are involved in the process. 
Since we consider $m_{DM} \gg m_{Z^\prime}$ and the DM production from thermal plasma 
  practically stops when $T$ becomes lower than $m_{DM}$ due to kinematics, 
  we can neglect $m_{Z^\prime}$ and $\Gamma_{Z^\prime}$ in Eq.~(\ref{DMSigma}) and 
  simplify the cross section formula to be 
\bea 
 \sigma(s) \simeq \frac{g_X^4}{48 \pi}  \frac{\sqrt{s (s-4 m_{DM}^2)}} {s^2} F(x_H).  
\label{DMSigma2}
\eea

\begin{figure}[t]
\begin{center}
{\includegraphics*[width=0.6\linewidth]{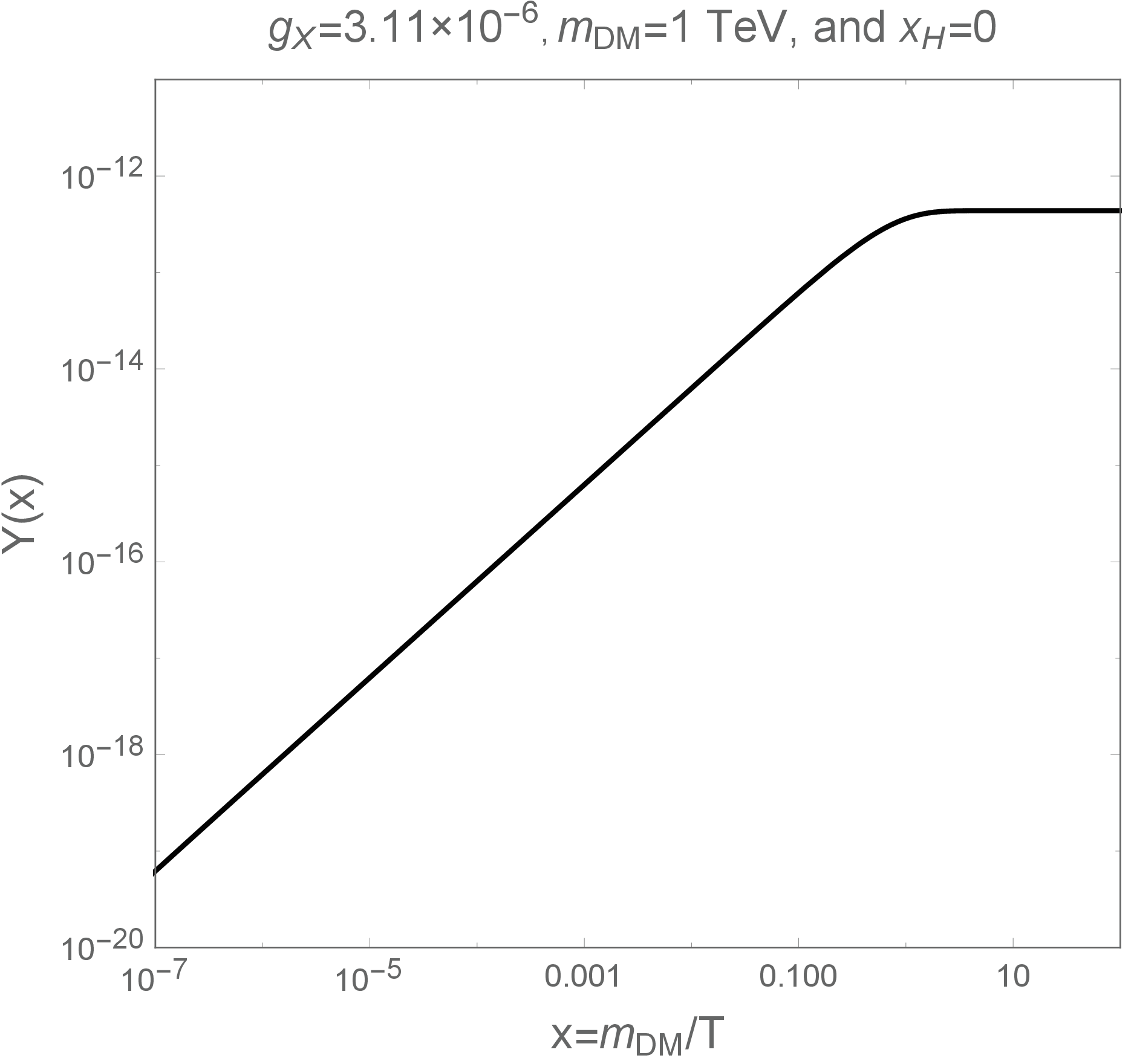}}
\caption{
Numerical solution of the Boltzmann equation for $m_{DM} \gg m_{Z^\prime}$. 
Here, we have set $g_X=3.11 \times 10^{-6}$, $m_{DM}=1$ TeV, $x_H=0$,   
   and $x_{RH}=10^{-10}$. 
The resultant $Y(\infty)$ reproduces the observed DM density of $\Omega_{DM} h^2=0.12$.   
}
\label{Fig:YA}
\end{center}
\end{figure}
  
Note that only three parameters, $m_{DM}$, $g_X$ and $x_H$, are involved in our analysis. 
Substituting Eq.~(\ref{DMSigma2}) into Eq.~(\ref{ThAvgSigma}) with fixed values of these parameters, 
  we numerically solve the Boltzmann equation of  Eq.~(\ref{Y:Boltzmann1}) 
  with the initial condition $Y(x_{RH})=0$ for $x_{RH} \ll 1$.  
Our result for $Y(x)$ is shown Fig.~\ref{Fig:YA}, 
   where we have set $g_X=3.11 \times 10^{-6}$, $m_{DM}=1$ TeV, $x_H=0$ (the $B-L$ model limit),     
   and $x_{RH}=10^{-10}$.    
For the parameter set, the resultant $Y(\infty) =4.36 \times 10^{-13}$ reproduces 
   the observed DM relic density $\Omega_{DM} h^2 =0.12$. 
We have checked that our result for the $B-L$ limit ($x_H=0$) is consistent with the one obtained in Ref.~\cite{Kaneta:2016vkq}.

In fact, it is easy to show that $Y(\infty)$ is independent of $x_{RH} \ll 1$  
   and $Y(\infty) \propto 1/m_{DM}$ so that $\Omega_{DM} h^2$ is independent of $m_{DM}$. 
Since $Y$ for the freeze-in DM never reaches $Y_{EQ}$ for $x \lesssim 1$ 
   because of its extremely weak interaction with the SM particles, 
   we can approximately express the Boltzmann equation of Eq.~(\ref{Y:Boltzmann1}) to be 
\bea 
\frac{d Y}{d x} \simeq \frac{s(m_{DM})}{H(m_{DM})} \,  \frac{\langle\sigma v_{\rm rel} \rangle}{x^2} \, Y_{EQ}^2
   \simeq  \frac{2.8}{g_*^{3/2}} \, m_{DM} \, M_P \,  \frac{\langle\sigma v_{\rm rel} \rangle}{x^2}  
\label{Y:Boltzmann2}
\eea 
for $x \lesssim 1$. 
For a given $\langle\sigma v_{\rm rel} \rangle$ as a function of $x$, 
  it is easy to solve the Boltzmann equation. 
The resultant $Y(\infty)$ is approximated as $Y(\infty) \simeq Y(x=1)$ 
  since the production of DM particles from the thermal plasma stops around $x\sim1$, 
  or equivalently, $T \sim m_{DM}$ due to kinematics.   
We can confirm this behavior in Fig.~\ref{Fig:YA}, where the numerical solution $Y(x)$ 
  quickly approaches $Y(\infty)$ at $x \sim 1$.   
For $x \lesssim 1$, we can approximate $\langle\sigma v_{\rm rel} \rangle$ by  
\bea 
  \langle\sigma v_{\rm rel} \rangle \simeq \frac{g_X^4}{384 \pi} \frac{x^2}{m_{DM}^2} F(x_H). 
\eea 
Substituting this into Eq.~(\ref{Y:Boltzmann2}), we integrate the Boltzmann equation 
  from $x_{RH} \ll1 $ to $x=1$ to obtain 
\bea
   Y(\infty) \simeq Y(x=1) \simeq 2.3 \times 10^{-3}  \, \frac{g_X^4}{g_\star^{3/2}} \, \frac{M_P}{m_{DM}}. 
\eea
Thus, we find $Y(\infty) \propto 1/m_{DM}$ and then 
\bea
  \Omega_{\rm DM} h^2  \simeq \frac{m_{DM} \, Y(x=1) \, s_0}{\rho_c/h^2} 
  \simeq 1.4 \times 10^{21}  \, \left( \frac{106.75}{g_\star} \right)^{3/2} g_X^4 \, F(x_H), 
\label{Omega-appx}
\eea
which is independent of $m_{DM}$. 
For the $B-L$ limit ($x_H=0$), we obtain $g_X\simeq 1.6 \times 10^{-6}$ to achieve the observed DM relic density 
  of $\Omega_{\rm DM} h^2=0.12$. 
Hence, this rough estimate leads to the $g_X$ value to be close to what we have obtained 
  by the numerical analysis, $g_X=3.11 \times 10^{-6}$.  
Considering that the cross section in Eq.~(\ref{DMSigma}) is proportional to $g_X^4 F(x_H)$ 
 (see also Eq.~(\ref{Omega-appx})),  
  we find that the observed DM relic density is reproduced by 
\bea 
    g_X = 3.11 \times 10^{-6} \left( \frac{F(0)}{F(x_H)} \right)^{1/4} 
 \label{g_X}   
\eea  
  for a general $x_H$ value.  

\begin{figure}[t]
\begin{center}
{\includegraphics*[width=0.6\linewidth]{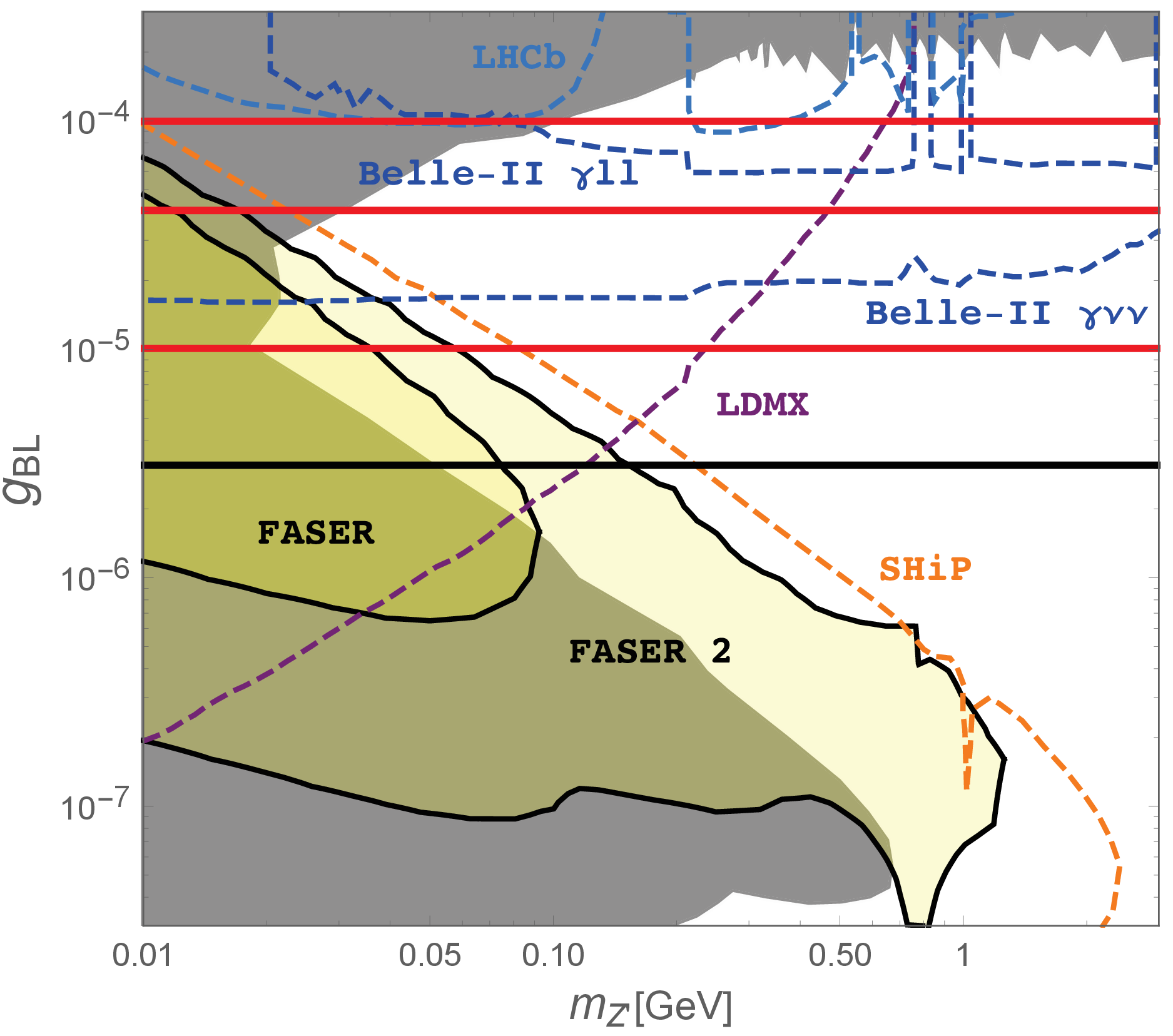}}
\caption{
Inferred $g_{BL}$ values to reproduce the observed DM relic density for various $|x_H|$ values 
   along with the search reach of various planned/proposed experiments at the Lifetime Frontier 
   and the current excluded region (gray shaded). 
The horizontal lines from top to bottom correspond to $|x_H|=900$, 150, 10 and 0 (the $B-L$ limit), respectively.     
}
\label{Fig:LF1}
\end{center}
\end{figure}

We now discuss how to test our scenario in the future experiments at the Lifetime Frontier. 
In order to reproduce the observed relic density for the RHN DM via the light $Z^\prime$-portal interaction, 
   the U(1)$_X$ gauge coupling is found to be very small as shown in Eq.~(\ref{g_X}).
This fact indicates that the $Z^\prime$ boson is long-lived. 
Such a long-lived particle can be explored at Lifetime Frontier experiments. 
The recently approved ForwArd Search Experiment (FASER) \cite{Feng:2017vli, Ariga:2018uku, Ariga:2019ufm} 
    plans its operation at the LHC Run-3 and its upgraded version (FASER 2) at the High-Luminosity LHC. 
The prospect for the $B-L$ gauge boson search at FASER is summarized in Ref.~\cite{Ariga:2018uku}.  
FASER 2 can search for a long-lived $Z^\prime$ boson 
   with mass in the range of 10 MeV$\lesssim m_{Z^\prime} \lesssim 1$ GeV
   for the $B-L$ gauge coupling in the range $10^{-8}  \lesssim g_{BL} \lesssim 10^{-4.5}$. 
The planned/proposed experiments, 
   such as Belle-II \cite{Dolan:2017osp}, LHCb \cite{Ilten:2015hya, Ilten:2016tkc}, SHiP \cite{Alekhin:2015byh}
   and LDMX \cite{Berlin:2018bsc}, 
   which will also search for a long-lived $Z^\prime$ boson,  
   will cover a parameter region complementary to FASER.

To obtain the prospect of our $Z^\prime$ boson search by the Lifetime Frontier experiments,  
   we need to interpret the analysis result for the $B-L$ gauge boson to the U(1)$_X$ model case. 
Since the results for $|x_H| \lesssim 1$ are expected to be very similar to the one for the $B-L$ case, 
   we are particularly interested in the hyper-charge oriented case of $|x_H| \gg 1$. 
From Table \ref{table1}, we see that the $Z^\prime$ boson coupling with the SM fermions 
   is controlled by $g_{BL}$ ($g_X$ for the $B-L$ limit of $x_H=0)$, 
   while it is controlled by $g_X |x_H|$ for $|x_H| \gg 1$. 
Hence, we express a correspondence between the $B-L$ gauge coupling ($g_{BL}$)
   and $g_X$ coupling such that 
\bea
     g_{BL}  \leftrightarrow g_X |x_H|. 
\eea
Eq.~(\ref{g_X}) leads to $g_X \simeq \frac{2.91 \times 10^{-6}}{\sqrt{|x_H|}}$ 
   for $|x_H| \gg 1$ for reproducing $\Omega_{DM} h^2 =0.12$.  
Therefore, the $g_{BL}$ value used in the analysis of the prospective search reach for the $B-L$ gauge boson 
   can be inferred to be  
\bea
     g_{BL}  \to g_X |x_H|  \simeq  2.91 \times 10^{-6}  \,\sqrt{|x_H|}.
\eea
In Fig.~\ref{Fig:LF1}, we show our results for the inferred $B-L$ gauge coupling 
   as a function of $m_{Z^\prime}$ to reproduce the observed DM relic density, 
   along with the search reach of various planned/proposed experiments at the Lifetime Frontier. 
The current excluded region from the combination of the searches for a long-lived particle
   and anomalous neutrino interactions is gray shaded (see Ref.~\cite{Bauer:2018onh} for details). 
The horizontal lines from top to bottom, along which $\Omega_{DM} h^2 =0.12$ is reproduced, 
   correspond to $|x_H|=900$, 150, 10 and 0 (the $B-L$ limit), respectively.  
The inferred $g_{BL}$ value shifts upward as $|x_H|$ increases. 
This result shows the impact of $x_H$ values on the future experiments. 
If a long-lived $Z^\prime$ boson is observed in the future, 
  we can determine $|x_H|$ and $m_{Z^\prime}$.

\begin{figure}[t]
\begin{center}
{\includegraphics*[width=0.6\linewidth]{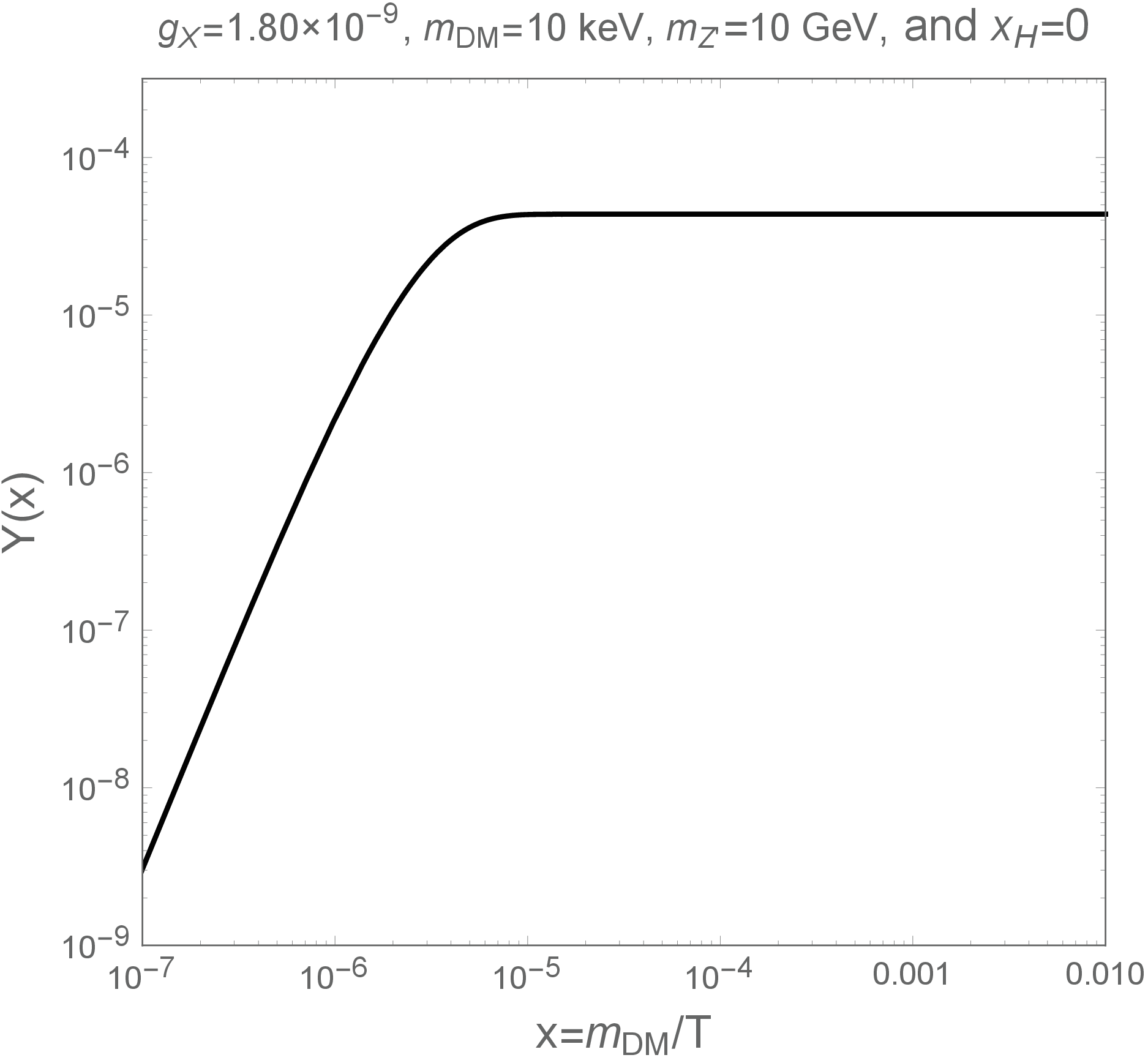}}
\caption{
Numerical solutions of the Boltzmann equation for $m_{DM} \ll m_{Z^\prime}$. 
Here, we have set $g_X=1.80 \times 10^{-9}$, $m_{DM}=10$ keV,  $m_{Z^\prime}=10$ GeV, 
   $x_H=0$, and $x_{RH}=10^{-10}$. 
The resultant $Y(\infty)$ reproduces the observed DM density of $\Omega_{DM} h^2=0.12$.   
}
\label{Fig:YB}
\end{center}
\end{figure}

\subsection{Case (ii): $m_{Z^\prime} \gg m_{DM}$}
\label{sec:3-2} 
We next consider the case $m_{Z^\prime} \gg m_{DM}$. 
Although the basic formulas that we employ in our analysis are same as in Case (i), 
   the RHN DM is dominantly produced through the $Z^\prime$ boson resonance in Case (ii).  
Note that for $m_{Z^\prime} > m_{DM}$, the DM pair creation/annihilation cross section of Eq.~(\ref{DMSigma}) 
   includes the resonance point at $s = m_{Z^\prime}^2$ for $s \geq 4 m_{DM}^2$.  
Since the gauge coupling is very small, we use the narrow width approximation,  
\bea
  \frac{ds}{(s-m_{Z^\prime}^2)^2 + m_{Z^\prime}^2 \Gamma_{Z^\prime}^2} 
  = ds \, \frac{\pi}{ m_{Z^\prime} \, \Gamma_{Z^\prime}}  \, \delta(s - m_{Z^\prime}^2), 
\eea
in calculating the thermal-averaged cross section, where the total decay width is given by\footnote{
$F(x_H)$ in the total decay width formula depends on $m_{Z^\prime}$ 
  since only the kinematically allowed final states are involved in the formula. 
For example, if a $Z^\prime$ boson is lighter than the top quark, Eq.~(\ref{F}) must be modified. 
However, $F(x_H) \gg 1$ is satisfied in our analysis, and our result 
  is almost independent  of $F(x_H)$.  
See Eq.~(\ref{DMSigma3}) and the discussion below. 
}
\bea
    \Gamma_{Z^\prime} = \frac{g_X^2}{24 \pi} \, m_{Z^\prime}  \left( F(x_H)  + 1  \right). 
\eea
We then obtain an analytic formula, 
\bea
  \langle\sigma v_{\rm rel} \rangle = \frac{g_X^4}{1536} \, \frac{x^5  \, m_{Z^\prime}^4}{m_{DM}^5 \Gamma_{Z^\prime} } 
    \, F(x_H) \, K_1\left( \frac{m_{Z^\prime}}{m_{DM}}x  \right)
    \simeq 
    \frac{\pi}{64} \, g_X^2 \, \frac{m_{Z^\prime}^3}{m_{DM}^5} \, x^5 \, K_1\left( \frac{m_{Z^\prime}}{m_{DM}}x  \right). 
 \label{DMSigma3}
\eea
Here, we have used $F(x_H) \gg 1$ in the last expression.

Four parameters, $m_{DM}$, $m_{Z^\prime}$, $g_X$ and $x_H$, are involved in our analysis. 
Using Eq.~(\ref{DMSigma3}) with fixed values of these parameters, 
  we numerically solve the Boltzmann equation of  Eq.~(\ref{Y:Boltzmann1}) 
  with the initial condition $Y(x_{RH})=0$ for $x_{RH} \ll 1$.  
Our result for $Y(x)$ is shown Fig.~\ref{Fig:YB}, 
   where we have set $g_X=1.80 \times 10^{-9}$, $m_{DM}=10$ keV,  $m_{Z^\prime}=10$ GeV, 
   $x_H=0$, and $x_{RH}=10^{-10}$.  
For the parameter set, the resultant $Y(\infty) =4.36 \times 10^{-5}$ reproduces 
   the observed DM relic density $\Omega_{DM} h^2 =0.12$. 
We have checked that our result for the $B-L$ limit ($x_H=0$) is consistent with the one obtained in Ref.~\cite{Kaneta:2016vkq}.

Also for Case (ii), we can derive an approximation formula for the DM relic density. 
We first notice that $K_1(y) \sim 1/y$ for $y \lesssim 1$ while $K_1(y) \propto e^{-y}$ for $y > 1$. 
Using this property of $K_1$, we may roughly approximate Eq.~(\ref{DMSigma3}) by
\bea
   \langle\sigma v_{\rm rel} \rangle 
    \sim    \frac{\pi}{64} \, g_X^2 \, \frac{m_{Z^\prime}^2}{m_{DM}^4} \, x^4 
 \label{DMSigma4}
\eea
  for $x \lesssim m_{DM}/m_{Z^\prime}$, 
  while $ \langle\sigma v_{\rm rel} \rangle \sim 0$ for $x > m_{DM}/m_{Z^\prime}$. 
 From the physics point of view, this formula looks reasonable, since for $x > m_{DM}/m_{Z^\prime}$, or equivalently  
  $T < m_{Z^\prime}$, the energy of the SM particles in the thermal plasma becomes too low 
  to create the $Z^\prime$ boson resonance and the DM creation rate drops significantly.  
Using this approximation, we integrate the Boltzmann equation from $x=x_{RH}$ to $x=m_{DM}/m_{Z^\prime}$ 
   with the initial condition $Y(x_{RH})=0$ to find 
\bea
  Y(\infty) \sim Y(x=m_{DM}/m_{Z^\prime}) \simeq 4.2 \times 10^{-6} \, g_X^2 \, \left(\frac{M_P}{m_{Z^\prime}} \right), 
\eea   
which is independent of $m_{DM}$.  
Using this formula, we find $g_X \sim 6.5 \times 10^{-12} \sqrt{\frac{m_{Z^\prime}}{m_{DM}}}$ 
   to obtain $\Omega_{DM} h^2=0.12$. 
Comparing this with our numerical result $g_X=1.80 \times 10^{-9}$ 
   to achieve $\Omega_{DM} h^2=0.12$ for $m_{DM}=10$ keV and $m_{Z^\prime}=10$ GeV,  
   we see that the rough estimate provides us with the right order of magnitude. 
In fact, we find from the numerical analysis that the relation of $g_X \propto  \sqrt{\frac{m_{Z^\prime}}{m_{DM}}}$  
   is satisfied to a very good accuracy. 
Through numerical analysis, we find that the observed DM relic density is reproduced by 
\bea
   g_X \simeq 1.80 \times 10^{-12} \, \sqrt{\frac{m_{Z^\prime}}{m_{DM}}}. 
\eea

\begin{figure}[t]
\begin{center}
{\includegraphics*[width=0.6\linewidth]{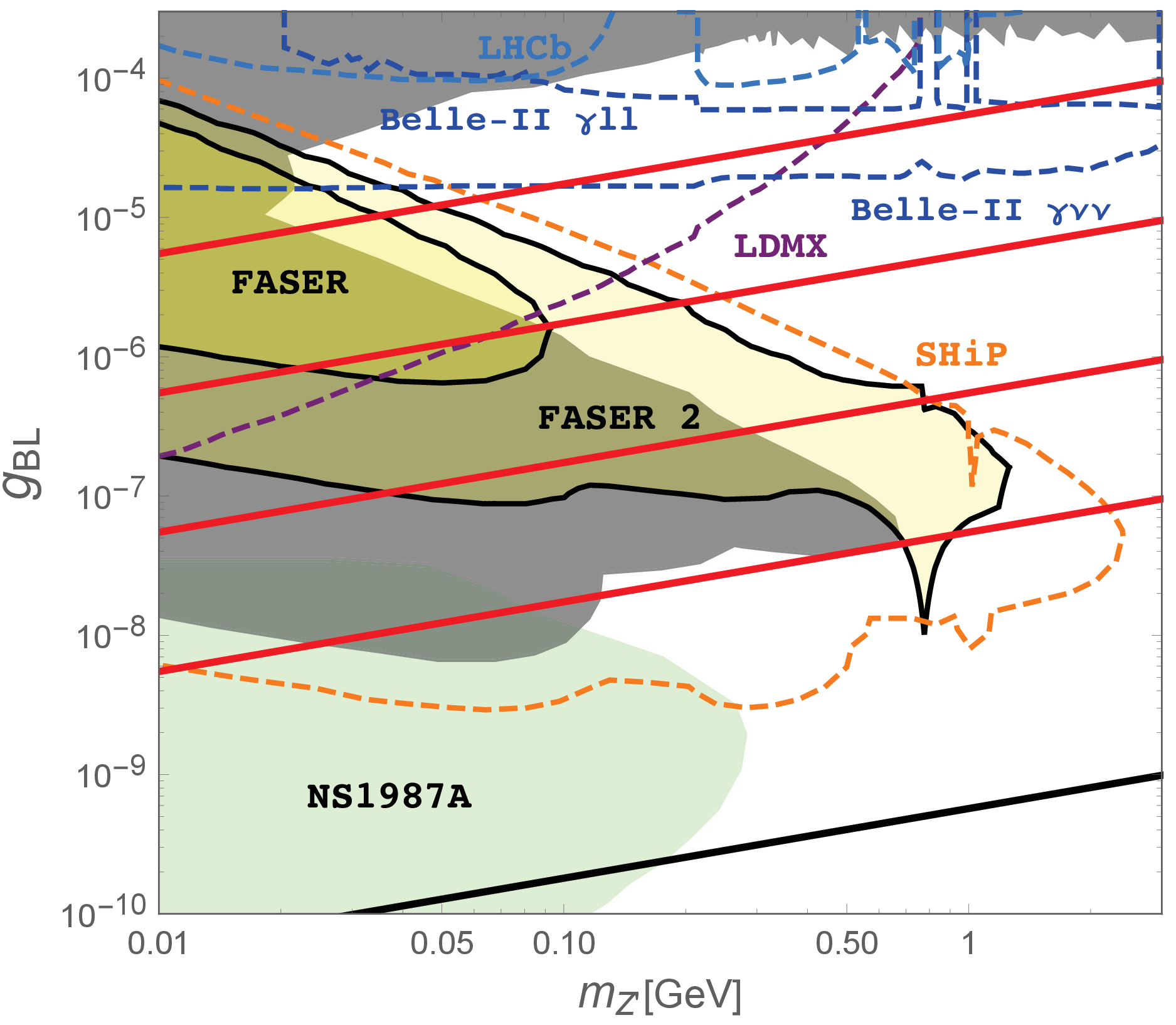}}
\caption{
Inferred $g_{BL}$ values to reproduce the observed DM relic density for various $|x_H|$ values 
   along with the search reach of various planned/proposed experiments at the Lifetime Frontier 
   and the current excluded region (gray shaded and light green shaded from the SN1987A observation). 
Here, we have set $m_{DM}=10$ keV. 
The diagonal lines from top to bottom, along which $\Omega_{DM} H^2 =0.12$ is reproduced, 
   correspond to $|x_H|=10^5$, $10^4$, $10^3$, $100$ and 0 (the $B-L$ limit), respectively.     
}
\label{Fig:LF2}
\end{center}
\end{figure}

To explore the prospect of the $Z^\prime$ boson search for Case (ii), 
   we now try to find a relation between $g_{BL}$  and $g_X$. 
Since the resultant $g_X$ is independent of $x_H$, 
   the $g_{BL}$ value used in the analysis  for the prospective search reach for the $B-L$ gauge boson 
   can be inferred to be
\bea
     g_{BL}  \to g_X |x_H|  \simeq  1.80 \times 10^{-12} \, \sqrt{\frac{m_{Z^\prime}}{m_{DM}}} \, |x_H|. 
\eea
In Fig.~\ref{Fig:LF2}, we show our results for the inferred $B-L$ gauge coupling 
   as a function of $m_{Z^\prime}$ to reproduce the observed DM relic density, 
   along with the search reach of various planned/proposed experiments at the Lifetime Frontier. 
The current excluded region from the combination of the searches for a long-lived particle
   and anomalous neutrino interactions is gray shaded (see Ref.~\cite{Bauer:2018onh} for details). 
The observation of Supernova 1987A (SN1987A) \cite{Hirata:1987hu, Bionta:1987qt} 
   excludes the green shaded region, 
   which causes an extra energy release for the supernova explosion 
   via $Z^\prime$ boson emissions \cite{Dent:2012mx, Kazanas:2014mca}.  
The diagonal lines from top to bottom, along which $\Omega_{DM} h^2 =0.12$ is reproduced, 
   correspond to $|x_H|=10^5$, $10^4$, $10^3$, $100$ and 0 (the $B-L$ limit), respectively.  
The inferred $g_{BL}$ value shifts upward as $|x_H|$ increases. 
If a long-lived $Z^\prime$ boson is observed in the future, 
  we can determine $|x_H|$ and $m_{Z^\prime}$.

\section{Conclusion and discussion}
\label{sec:4}
The minimal gauged U(1)$_X$ extension of the SM is a simple, well-motivated framework 
   to incorporate the neutrino masses in the SM, 
   where the U(1)$_X$ charge of a field is defined as a linear combination of its hyper-charge and $B-L$ charge. 
In addition to the U(1)$_X$ gauge boson ($Z^\prime$), three right-handed neutrinos and 
   one U(1)$_X$ Higgs field are introduced.    
The anomalies related to the U(1)$_X$ gauge interaction are all canceled 
   in the presence of three right-handed neutrinos. 
In addition to all the Yukawa couplings for charged leptons and quarks in the SM, 
   the U(1)$_X$ gauge symmetry also allows new types of Yukawa couplings for neutrinos: 
   Dirac-type neutrino Yukawa couplings between the SM left-handed neutrinos and the new right-handed neutrinos,  
   and Majorana-type Yukawa couplings for the right-handed neutrinos. 
With the breaking of the electroweak and U(1)$_X$ symmetries, these Yukawa couplings generate 
   Dirac-type and Majorana-type masses for neutrinos and then the seesaw mechanism 
   induces the desired light neutrino masses naturally. 
This minimal U(1)$_X$ model is supplemented with a dark matter candidate 
   by introducing a $Z_2$ symmetry such that one right-handed neutrino being a unique $Z_2$-odd field 
   serves as the dark matter in the universe. 
The two remaining right-handed neutrinos are involved in the minimal seesaw mechanism 
   and reproduce the neutrino oscillation data with one massless neutrino eigenstate.

We have investigated the phenomenology of this Majorana fermion dark matter
   which communicates with the SM sector through the $Z^\prime$ gauge boson. 
Motivated by the future Lifetime Frontier experiments, we have focused on 
   the parameter space where the dark matter particle very weakly couples 
   to the light $Z^\prime$ boson. 
In this case, the $Z^\prime$ boson is long-lived and its displaced vertex signature 
   can be explored by the future experiments if its mass lies in the range of 10 MeV$-$a few GeV.
Since the U(1)$_X$ gauge interaction is very weak, the dark matter particle had never been
   in thermal equilibrium with the SM particles, and the dark matter relic density is determined 
   through the freeze-in mechanism. 
For the two cases, $m_{DM} \gg m_{Z^\prime}$ and $m_{DM} \ll m_{Z^\prime}$, 
   we identify the model parameter regions to reproduce the observed DM density
   of $\Omega_{DM} h^2=0.12$.
We have also discussed how our scenario can be tested by various future Lifetime Frontier experiments. 
In particular, we have found a measurable impact for $|x_H| \gg 1$ on the experiments. 
The U(1)$_X$ extension of the $B-L$ model with a large $|x_H|$ dramatically alters 
   the parameter region to be explored by the future experiments 
   compared to that previously obtained for the $B-L$ model ($x_H=0$ limit in our framework).

Finally, we extract the U(1)$_X$ symmetry breaking scale 
   which corresponds to the parameter regions shown in Figs.~\ref{Fig:LF1} and \ref{Fig:LF2}. 
In Case (i), we have obtained $g_X \simeq \frac{2.91 \times 10^{-6}}{\sqrt{|x_H|}}$ 
   for $|x_H| \gg 1$ to reproduce $\Omega_{DM} h^2 =0.12$.  
 From Eq.~(\ref{Mass}) this leads to 
\bea
   v_X = \frac{m_{Z^\prime}}{2 g_X} \simeq 1.7 \times 10^5 \, [{\rm GeV}] \, \sqrt{|x_H|} \, 
    \left(  \frac{m_{Z^\prime}}{1 \, {\rm GeV}}\right)
\eea   
for $|x_H| \gg 1$. 
For Case (ii), we find 
\bea
   v_X = \frac{m_{Z^\prime}}{2 g_X} \simeq  2.8 \times 10^{11} \, \sqrt{m_{DM} \, m_{Z^\prime}}
      \simeq 8.8 \times 10^8 \, [{\rm GeV}] \, 
      \sqrt{\frac{m_{DM}}{10 \,  {\rm keV}}} \,  \sqrt{\frac{m_{Z^\prime}}{1 \,  {\rm GeV}}}. 
\eea
These $v_X$ values indicate from Eq.~(\ref{Mass}) 
   that the Majorana Yukawa coupling ($Y_N$) is also very small. 
On the other hand, we may set $Y_N^{1,2}$ to have a nearly degenerate spectrum of 
   $m_N^1 \simeq m_N^2 \gtrsim 1$ TeV,  
   which allows for a successful leptogenesis through an enhancement of 
   the $CP$-violating parameter (resonant leptogenesis \cite{Flanz:1996fb, Pilaftsis:1997jf, Pilaftsis:2003gt}).\footnote{
See Ref.~\cite{Caputo:2018zky} for a scenario similar to Case (ii), but a successful leptogenesis is realized 
   through $CP$-violating right-handed neutrino oscillations.}

\section*{Acknowledgments}
The authors N.O.~and S.O.~would like to thank the Particle Theory Group of the University of Delaware 
  for hospitality during their visit.    
This work is supported in part by 
  the United States Department of Energy Grants DE-SC-0012447 (N.O.) and DE-SC0013880 (Q.S.), 
  and the M. Hildred Blewett Fellowship of the American Physical Society, www.aps.org (S.O.).

\bibliographystyle{utphysII}
\bibliography{References}

\end{document}